\documentclass[12pt]{article}
\usepackage{epsf}
\newcommand{\be}{\begin{equation}}
\newcommand{\ee}{\end{equation}}
\begin{document}

\title{Absence of higher order corrections to the non-Abelian topological mass
term}

\author{F. T. Brandt$^\dagger$, Ashok Das$^\ddagger$ and J. Frenkel$^\dagger$ 
\\ \\
$^\dagger$Instituto de F\'{\i}sica,
Universidade de S\~ao Paulo\\
S\~ao Paulo, SP 05315-970, BRAZIL\\
$^\ddagger$Department of Physics and Astronomy\\
University of Rochester\\
Rochester, NY 14627-0171, USA}

\maketitle

\begin{abstract}

We study the Yang-Mills-Chern-Simons theory systematically in an
effort to generalize the Coleman-Hill result to the non-Abelian
case. We show that, while the Chern-Simons coefficient is in general
gauge dependent in a non-Abelian theory, it takes on a physical
meaning in the axial gauge. Using the non-Abelian Ward identities as
well as the analyticity of the amplitudes in the momentum variables,
we show that, in the axial gauge, the Chern-Simons coefficient does
not receive any quantum correction beyond one loop. This allows us to
deduce that the ratio ${4\pi m\over g^{2}}$ is unrenormalized, in a
non-Abelian theory, beyond one loop in any infrared safe gauge. This
is the appropriate generalization of the Coleman-Hill result to
non-Abelian theories. Various other interesting properties of the
theory are also discussed.

\end{abstract}
\vfill\eject

\section{Introduction:}

It is well known by now that in odd space-time dimensions, one can add
a topological term to the Lagrangian density of a gauge field, in
addition to the usual Yang-Mills (Maxwell) term. Such a term is known as the
Chern-Simons term and a theory with a Chern-Simons term is
conventionally called a Chern-Simons theory {\cite{1,2}}.
In $2+1$ dimensions, for
example, the Chern-Simons term, in a $SU(N)$ gauge theory, has the form
\[
{\cal L}_{CS} = {m\over 2}\,\epsilon^{\mu\nu\lambda}\,A_{\mu}^{a}
(\partial_{\nu}A_{\lambda}^{a} + {g\over 3} f^{abc}
A_{\nu}^{b}A_{\lambda}^{c})
\]
where $g$ represents the gauge coupling and $f^{abc}$ stand for the
structure constants of the group. The parameter $m$ is known as the
Chern-Simons coefficient (at the tree level) and has the dimensions of
mass.  In a theory,
with a Yang-Mills term for the gauge fields, it can be shown that the
Chern-Simons term provides a gauge invariant mass for the gauge
fields. Such a mass term is absolutely crucial in the perturbative
study of a pure Yang-Mills-Chern-Simons gauge theory, since without this
term, the infrared divergences in $2+1$ dimensions are so severe that
a perturbative expansion cannot be defined {\cite{2,3}}.

The Chern-Simons term violates discrete symmetries like $P$ and $T$
(although it respects $CPT$).
In a gauge theory with (matter) interactions  which
violate these symmetries, it is expected that a Chern-Simons term will
be generated at the quantum level, even if one is not present at the
tree level. Thus, for example, the mass term for a fermion, in $2+1$
dimensions, is known to violate these symmetries and, correspondingly,
it is known that a massive fermion interacting with gauge fields
generates a Chern-Simons term at the one loop level  {\cite{2}}.
Surprisingly,
however, it was noted, through explicit calculations, that even though
a Chern-Simons term is generated at one loop, there is no radiative
correction to the Chern-Simons coefficient at the two loop level, either in
the Abelian or in the non-Abelian theory {\cite{4,5}}.
This peculiarity was
explained, for the Abelian theory, by Coleman and Hill, who showed
that, in $2+1$ dimensional QED with or without a tree level
Chern-Simons term, the Chern-Simons coefficient does not receive any
contribution beyond one loop at zero temperature {\cite{6}}.
The proof of this
result is quite elegant and  essentially
uses two key assumptions, namely, i) the Abelian Ward identity and,
ii) the analyticity of the amplitudes in the momentum
(energy-momentum) variables. This result holds, in an Abelian theory,
whenever these assumptions are valid, but not otherwise.

While the Coleman-Hill result explains the peculiarity of the explicit two
loop calculation in the Abelian theory, it says nothing about the
outcome of the 
calculation in the non-Abelian theory. There are, in fact, several
difficulties one faces in trying to extend the Coleman-Hill result to
non-Abelian theories. First, unlike in an Abelian theory at zero
temperature, without a tree level Chern-Simons term, infrared
divergences may be too severe (as we have already mentioned). Second,
even with a tree level Chern-Simons term 
in the non-Abelian theory, an arbitrary gauge choice may introduce
spurious infrared divergences and, therefore, one must carefully
choose an infrared safe gauge {\cite{2,7}} (again, this is
not a  problem in the Abelian theory). Finally, 
in the non-Abelian theory, the Chern-Simons coefficient is, in
general, gauge dependent (in an Abelian theory, this coefficient is
gauge independent). Therefore, an attempt to naively  generalize the
Coleman-Hill result is meaningless. On the other hand, it is known
that, in a non-Abelian theory, the ratio ${4\pi m\over g^{2}}$ is
gauge independent and has a physical significance. 
Consequently, it makes sense to try and show that it is this ratio 
which gets no contribution beyond one loop in a non-Abelian theory.

In this paper, we show that this expectation, indeed, holds. In
particular, we show, much like the Coleman-Hill result in the Abelian
theory, that if, i) the Ward identities of the non-Abelian theory hold
and, ii) the amplitudes are analytic in the momentum (energy-momentum)
variables, then the ratio ${4\pi m\over g^{2}}$ does not receive any
quantum correction beyond one loop. The non-Abelian theory is clearly 
much more complicated than the Abelian counterpart and we prove our
result by working in the axial gauge, which is an infrared safe
gauge. It is, of course, known that the Ward identities in the axial
gauge are much simpler, but we show that, in this gauge, the
Chern-Simons coefficient takes on a physical significance, although it
is gauge dependent in general. In fact, we show that the Chern-Simons
coefficient, in the axial gauge, receives no quantum correction beyond
one loop and this allows us to deduce that the ratio 
${4\pi m\over g^{2}}$ is unaffected beyond one loop. A brief account
of our main result has already been published {\cite{8}}
and here we describe the details of our work along with many other
interesting features of the analysis.

The organization of our paper is as follows. In section {\bf 2}, we
analyze the Yang-Mills-Chern-Simons theory in a covariant gauge and show,
using a Nielsen-like identity {\cite{9,10}},
that the Chern-Simons coefficient is, in
general, gauge dependent. In section {\bf 3}, we define the theory in
the axial gauge and discuss some of the special features of this
gauge choice. From the Ward identities, in this gauge, we obtain a diagrammatic
representation for the Chern-Simons coefficient, which is quite useful
in an all order proof. We also show that, in this gauge, the Chern-Simons
coefficient takes on a physical meaning and derive a Nielsen-like
identity to show that it is independent of $n^{\mu}$, the 
choice of direction in the axial gauge. In section {\bf 4}, we explicitly
evaluate the Chern-Simons coefficient at one loop and show that it is
independent of $n^{\mu}$ as is required from the Nielsen-like
identity described in section {\bf 3}. We compare our calculation with
that in the Landau gauge {\cite{7}} to bring out the
gauge  independent nature of
the ratio ${4\pi m\over g^{2}}$. We also present an interpolating
gauge that interpolates between the infrared safe Landau and axial
gauges. In section {\bf 5} we prove the main result of our paper,
namely that, with the assumptions of BRST invariance and analyticity
of the amplitudes, the Chern-Simons coefficient has no quantum
correction beyond one loop in the axial gauge. We deduce from this that,
in any infrared safe gauge, the ratio ${4\pi m\over g^{2}}$ receives
no radiative correction beyond one loop. In section {\bf 6}, we study
the pure Chern-Simons theory (without a Yang-Mills term) and show that it
has an additional vector supersymmetry in the axial gauge (much like
the one in the Landau gauge). The Ward
identities following from this, together with the usual Ward
identities show that this is a free theory. We present a brief
conclusion in section {\bf 7}. 

\section{Gauge Dependence:}

Let us consider the Yang-Mills-Chern-Simons theory in $2+1$ dimensions
described  by the Lagrangian density {\cite{2,7,11,12,13}}
\begin{equation}
{\cal L}_{inv} = {1\over 2} {\rm tr}\, F_{\mu\nu}F^{\mu\nu} - m\,{\rm
tr}\,\epsilon^{\mu\nu\lambda} A_{\mu}(\partial_{\nu}A_{\lambda} +
{2\,g\over 3}A_{\nu}A_{\lambda})\label{1}
\end{equation}
where we have chosen, for simplicity, the Chern-Simons mass $m$ to be positive.
The gauge field belongs to a matrix representation of $SU(N)$,
\[
A_{\mu} = A_{\mu}^{a}T^{a}
\]
with the generators of the group assumed to have the normalization
%%% why the minus sign in the normalization
\[
{\rm tr}\, T^{a}T^{b} = -{1\over 2} \delta^{ab}
\]
and
\[
F_{\mu\nu} = \partial_{\mu}A_{\nu} - \partial_{\nu}A_{\mu} + g
[A_{\mu},A_{\nu}].
\]
This is a self-interacting theory and one can, of course, add to it
interacting matter fields. However, we would restrict ourselves,
for simplicity, to the theory described by Eq. (\ref{1}), which can be
written with explicit internal symmetry indices as
\begin{equation}
{\cal L}_{inv} = - {1\over 4} F^{\mu\nu,a}F_{\mu\nu}^{a} + {m\over
2}\epsilon^{\mu\nu\lambda} A_{\mu}^{a} (\partial_{\nu}A_{\lambda}^{a}
+ {g\over 3}f^{abc}A_{\nu}^{b}A_{\lambda}^{c})\label{2}
\end{equation}

The Lagrangian density, in Eq. (\ref{1}), is invariant under the
infinitesimal $SU(N)$ gauge transformations of the form
\[
\delta A_{\mu}(x) = D_{\mu}\epsilon(x) = \partial_{\mu}\epsilon(x) + g
[A_{\mu},\epsilon ]
\]
where $\epsilon(x)$ is an infinitesimal matrix valued transformation
parameter. On the other hand, under a finite gauge transformation
\[
A_{\mu} \rightarrow U^{-1}A_{\mu}U - {i\over g} U^{-1}\partial_{\mu}U
\]
the Lagrangian density changes by a total divergence (it is the Chern-Simons
term that is not invariant), so that the action changes by a constant
\begin{equation}
S_{inv} = \int d^{3}x\,{\cal L}_{inv} \rightarrow S_{inv} + {4\pi m\over
g^{2}}\,2\pi W\label{2'}
\end{equation}
where
\begin{equation}
W = {1\over 24\pi^{2}} \int d^{3}x\,\epsilon^{\mu\nu\lambda}\,{\rm
tr}U^{-1}\partial_{\mu}U U^{-1}\partial_{\nu}U
U^{-1}\partial_{\lambda}U
\end{equation}
is an integer, known as the winding number of the gauge transformation,
and classifies the gauge transformations into topologically distinct
classes. When the winding number vanishes, the gauge transformations
are conventionally known as {\it small gauge} transformations,
while non zero winding numbers lead to {\it large gauge}
transformations. It is clear from Eq. (\ref{2'}) that under a {\it
small gauge} transformation, the action is invariant, while, under a
{\it large gauge} transformation, the action changes by a
constant. In the path integral approach, it is quite clear that even
though there is a shift in the action under a {\it large gauge}
transformation, the generating functional is invariant provided
\begin{equation}
{4\pi m\over g^{2}} = n\label{3}
\end{equation}
where $n$ is a positive integer (because of our choice $m>0$).

It is well known that the coefficient of the Chern-Simons term (which
is $m$ in the tree level), in an
Abelian theory, is a gauge independent quantity. It is related to the
physically meaningful statistics parameter and, in fact, it is this
coefficient which does not receive quantum corrections beyond one-loop
(provided certain assumptions are valid) according to the Coleman-Hill
result. In trying to extend this result to non-Abelian
theories, one of the challenges we face, as mentioned in the
introduction, is that the Chern-Simons coefficient is,
in general, a gauge dependent quantity in a non-Abelian theory. This
is best seen from the following analysis involving a Nielsen-like
identity  {\cite{9,10}}.

Let us analyze the Chern-Simons theory in a general covariant
gauge. Thus, adding a
gauge fixing and ghost Lagrangian density of the form
\begin{eqnarray}
{\cal L}_{gf} + {\cal L}_{ghost} & = & -{1\over 2\xi}
(\partial_{\mu}A^{\mu,a})^{2} + \partial^{\mu}\overline{c}^{a}
D_{\mu}c^{a}\nonumber\\
\noalign{\kern 4pt}%
 & = & {\xi\over 2} F^{a}F^{a} - F^{a}(\partial_{\mu}A^{\mu,a}) +
\partial^{\mu}\overline{c}^{a} D_{\mu}c^{a}
\end{eqnarray}
we can write the total Lagrangian density, in this gauge, to be
\begin{equation}
{\cal L} = {\cal L}_{inv} + {\cal L}_{gf} + {\cal L}_{ghost}
\end{equation}
We note that we have introduced an auxiliary field, $F^{a}$, to write the gauge
fixing term, which helps close the algebra of the BRST charges off-shell.
From the BRST identities for the theory, in this gauge, one knows that
the gauge fixing parameter, $\xi$, is not renormalized so that we can
parameterize the two point function of the full theory as
\begin{equation}
\begin{array}{ll}
\Pi^{\mu\nu,ab}(p) = \delta^{ab}& \left[\right. (p^{\mu}p^{\nu}
-\eta^{\mu\nu}p^{2})(1+\Pi_{1}(p))  \\ & \\
& +i\,m\epsilon^{\mu\nu\lambda}p_{\lambda}(1+\Pi_{2}(p)) - 
\displaystyle{1\over \xi} p^{\mu}p^{\nu}\left.\right]
\end{array}
\end{equation}
Here, $\Pi_{1}(p)$ and $\Pi_{2}(p)$ represent, respectively, the
radiative corrections to the parity conserving transverse part and the
parity violating part of the two point function. It is worth noting from this
that the Chern-Simons coefficient, at any order, can be obtained from the
two point function as
\begin{equation}
\delta^{ab}\overline{\Pi}_{2}(0) = \delta^{ab}(1+\Pi_{2}(0)) = {1\over
6im}\,\epsilon_{\mu\nu\lambda}\,\left.{\partial\over \partial
p_{\lambda}}\Pi^{\mu\nu,ab}(p)\right|_{p=0}\label{3'}
\end{equation}
We note that it is $m\overline{\Pi}_{2}(0)$ which represents the complete
Chern-Simons coefficient, with $m\Pi_{2}(0)$ representing the part
coming from quantum corrections. Let us also note here that, in this
gauge, the tree level propagator for the gauge field has the form
\begin{equation}
D_{\mu\nu}^{(0) ab}(p) = \delta^{ab}\left[{1\over
p^{2}-m^{2}}\left\{\left(\eta_{\mu\nu}-{p_{\mu}p_{\nu}\over
p^{2}}\right) + im\,\epsilon_{\mu\nu\lambda}\,{p^{\lambda}\over
p^{2}}\right\} + {\xi p_{\mu}p_{\nu}\over (p^{2})^{2}}\right]
\end{equation}

To study the gauge dependence of the Chern-Simons coefficient, let us
add to the Lagrangian the following source terms
\begin{equation}
{\cal L}_{total} = {\cal L} + {\cal L}_{source}
\end{equation}
where
\begin{eqnarray}
{\cal L}_{source} & = & J^{\mu,a}A_{\mu}^{a} + J^{a}F^{a} +
i(\overline{\eta}^{a}c^{a} - \overline{c}^{a}\eta^{a}) + K^{\mu,a}
D_{\mu}c^{a}\nonumber\\
\noalign{\kern 4pt}%
 &  & \quad + L^{a}(-{1\over 2}f^{abc}c^{b}c^{c}) + {1\over
 2}HF^{a}\overline{c}^{a}\label{3a}
\end{eqnarray}
Here, all  the sources are the standard ones, introduced to derive and study
BRST identities, except for the last term whose role would become
clear shortly.

We note that, under a BRST transformation ($\omega$ is a space-time
independent anti-commuting parameter),
\begin{eqnarray}
\delta A_{\mu}^{a} & = & \omega D_{\mu}c^{a}\nonumber\\
\noalign{\kern 4pt}%
\delta c^{a} & = & -{\omega\over 2}\,f^{abc}c^{b}c^{c}\nonumber\\
\noalign{\kern 4pt}%
\delta \overline{c}^{a} & = & - \omega F^{a}\nonumber\\
\noalign{\kern 4pt}%
\delta F^{a} & = & 0\label{4}
\end{eqnarray}
the source terms are not invariant although ${\cal L}$ is. In
fact, we obtain
\begin{equation}
\delta {\cal L}_{source} = \omega \left[J^{\mu,a}(D_{\mu}c^{a})
-i\overline{\eta}^{a} (-{1\over 2}f^{abc}c^{b}c^{c}) + iF^{a}\eta^{a}
+ {1\over 2} HF^{a}F^{a}\right]
\end{equation}
Making a field redefinition inside the path integral which coincides
with a BRST transformation, then, we obtain from the invariance of the
generating functional
\[
Z = e^{iW} = \int {\cal D}A_{\mu}^{a} {\cal D}F^{a} {\cal
D}\overline{c}^{a} {\cal D}c^{a}\, e^{i\int d^{3}x\,{\cal L}_{total}}
\]
the master identity
\begin{eqnarray}
{\partial W\over \partial\xi} & = & \int
d^{3}x\,d^{3}y\left(J^{\mu,a}(x){\delta^{2}W\over \delta
K^{\mu,a}(x)\delta H(y)} - i\overline{\eta}^{a}(x){\delta^{2}W\over
\delta L^{a}(x)\delta H(y)}\right.\nonumber\\
\noalign{\kern 4pt}%
 &  & \qquad\qquad\qquad \left. - i{\delta^{2}W\over \delta J^{a}(x)\delta
H(y)}\eta^{a}(x)\right)
\end{eqnarray}
In other words, this identity allows us to study the gauge dependence
of the effective action.

We can now make a Legendre transformation (with respect to the usual
sources $J^{\mu,a},J^{a},\eta^{a},\overline{\eta}^{a}$) and go to the
effective action and the identity above takes the form
\begin{eqnarray}
{\partial\Gamma\over \partial\xi} & = & - \int
d^{3}x\,d^{3}y\left({\delta\Gamma\over \delta
A_{\mu}^{a}(x)}{\delta^{2}\Gamma\over \delta K^{\mu,a}(x)\delta
H(y)} + {\delta\Gamma\over \delta c^{a}(x)}{\delta^{2}\Gamma\over \delta
L^{a}(x)\delta H(y)}\right.\nonumber\\
\noalign{\kern 4pt}%
 &  & \qquad\qquad\qquad \left. - {\delta F^{a}(x)\over\delta
H(y)}{\delta\Gamma\over \delta \overline{c}^{a}(x)}\right)
\end{eqnarray}
This identity describes the gauge dependence of the effective action
and we can derive the gauge dependence of any 1PI amplitude from
this. In particular, we note that (see Eq. (\ref{3'}))
\begin{eqnarray}
{\partial\overline{\Pi}_{2}(0)\over \partial\xi} =  \left.{1\over
6\, im\,(N^{2}-1)}\,\epsilon_{\mu\nu\rho}{\partial\over
\partial\xi}{\partial\over \partial p_{\rho}}
{\delta^{2}\Gamma\over \delta A_{\mu}^{a}(p)\delta
A_{\nu}^{a}(-p)}\right|_{p=0}\nonumber\\
\noalign{\kern 4pt}%
  =  
- {1\over 6\, im\, (N^{2}-1)}\,\epsilon_{\mu\nu\rho} {\partial\over
 \partial p_{\rho}}\left[{\delta^{3}\Gamma\over \delta
 A_{\mu}^{a}(p)\delta A_{\nu}^{a}(-p)\delta
 A_{\lambda}^{b}(0)}{\delta^{2}\Gamma\over \delta
 K^{\lambda,b}(0)\delta H(0)}\right.\nonumber  \\
\noalign{\kern 4pt}%
    \qquad + {\delta^{2}\Gamma\over \delta
 A_{\mu}^{a}(p)\delta A_{\lambda}^{b}(-p)}{\delta^{3}\Gamma\over
 \delta A_{\nu}^{a}(-p)\delta K^{\lambda,b}(p)\delta H(0)}\nonumber\\
\noalign{\kern 4pt}%
     \qquad \left.\left.+{\delta^{2}\Gamma\over \delta
 A_{\nu}^{a}(-p)\delta A_{\lambda}^{b}(p)}{\delta^{3}\Gamma\over
 \delta A_{\mu}^{a}(p)\delta K^{\lambda,b}(-p)\delta
 H(0)}\right]\right|_{p=0}\label{4a}
\end{eqnarray}
Here, we are supposed to also understand that all fields are set to
zero after evaluating the functional derivatives.

\begin{figure}[t!]
    \epsfbox{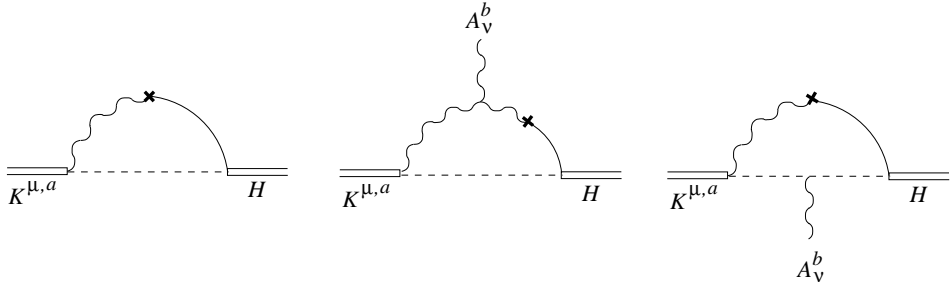}
\caption{One loop diagrams that can lead to a mixing of the
  sources. The wavy lines represent gauge fields, the solid line, the
  auxiliary field $F^a$, and the dashed lines stand for ghosts.}
\label{fig1}
\end{figure}

There are several things to note from this. First, there are no tree
level mixing terms of the forms $K^{\mu,a}H, K^{\mu,a}A_{\nu}^{b}H$ so
that the gauge dependence of the Chern-Simons coefficient can only arise from
radiative corrections. One can explicitly check at one loop level and
argue from symmetry arguments that radiative corrections cannot
generate a vertex of the form $K^{\mu,a}H$ (for such a vertex, the
colour index cannot be saturated). Consequently, the first
term on the right hand side of Eq. (\ref{4a}) does not contribute. At one
loop, a vertex of the form $K^{\mu,a}(p)A_{\nu}^{b}(-p)H(0)$ is
already generated (see Fig. 1). Therefore, let us parameterize such a
vertex as,
\begin{equation}
\left.{\delta^{3}\Gamma\over \delta K^{\mu,a}(p)\delta
A_{\nu}^{b}(-p)\delta H(0)}\right| =
\delta^{ab}\left[\delta_{\mu}^{\nu}A(p) + p_{\mu}p^{\nu}B(p) +
i\epsilon_{\mu}^{\;\nu\lambda}p_{\lambda}C(p)\right]
\end{equation}
Substituting this into the identity (\ref{4a}), we obtain,
\begin{equation}
{\partial\overline{\Pi}_{2}(0)\over \partial \xi} = 2
(1+\Pi_{2}(0))A(0)
\end{equation}
The right hand side can be evaluated order by order and, in general,
is not zero showing that the Chern-Simons coefficient, in a non-Abelian theory
is, in general, gauge dependent. We also note that $A(0)$ is obtained
from the vertex $K^{\mu,a}A_{\nu}^{b}H$ with all external momenta equal
to zero. Consequently, this has severe infrared divergences and unless
an infrared safe gauge, like the Landau gauge, is chosen, 
the identities cannot even be
satisfied. 

\section{Axial gauge:}

In the previous section, we saw that the Chern-Simons coefficient is, in
general, gauge dependent. Therefore, this naturally raises the question as to
whether the Coleman-Hill result can even be meaningfully generalized
to  non-Abelian
theories and if so, in what manner. In this section, we will show
that, in the axial gauge, the Chern-Simons coefficient has a physical
significance and, therefore, this is possibly the appropriate gauge in which to
consider a generalization of the Coleman-Hill analysis.

Let us consider a general axial gauge {\cite{14}} described by a
gauge fixing and ghost Lagrangian density of the form
\begin{eqnarray}
{\cal L}_{gf} + {\cal L}_{ghost} & = & - {1\over 2\xi}
(n^{\mu}A_{\mu}^{a})^{2} -
\overline{c}^{a}n^{\mu}(D_{\mu}c^{a})\nonumber\\
\noalign{\kern 4pt}%
 & = & {\xi\over 2} F^{a}F^{a} - F^{a}(n^{\mu}A_{\mu}^{a}) -
 \overline{c}^{a}n^{\mu}(D_{\mu}c^{a})\label{5}
\end{eqnarray}
Here, $n^{\mu}$ represents an  arbitrary direction. The theory
described by
\begin{equation}
{\cal L} = {\cal L}_{inv} + {\cal L}_{gf} + {\cal L}_{ghost}\label{6} 
\end{equation}
is infrared divergent in $2+1$ dimensions, unless $\xi = 0$ and we
will study the
theory in such a limiting gauge. For $\xi = 0$, $n^{2}=
n^{\mu}n_{\mu} = 0$ defines the light-cone gauge, while 
$n^{2}\geq 0$ leads to the time-like axial gauge and so on. 

The tree level propagator of the gauge field for an arbitrary gauge
fixing parameter $\xi$ is given by
\begin{eqnarray}
D^{(0) ab}_{\mu\nu} (p) & = & {\delta^{ab}\over
p^{2}-m^{2}}\left[\eta_{\mu\nu} - {n_{\mu}p_{\nu}+n_{\nu}p_{\mu}\over
(n\cdot p)} + {n^{2}p_{\mu}p_{\nu}\over (n\cdot p)^{2}} +
im\epsilon_{\mu\nu\lambda} {n^{\lambda}\over (n\cdot
p)}\right]\nonumber\\
\noalign{\kern 4pt}%
 &  & \qquad\qquad\qquad +
{\delta^{ab}\xi p_{\mu}p_{\nu}\over (n\cdot p)^{2}}
\end{eqnarray}
From this, we obtain the tree level propagator, in the axial gauge
($\xi=0$), to be
\begin{equation}
D^{(0) ab}_{\mu\nu}(p) =  {\delta^{ab}\over
p^{2}-m^{2}}\left[\eta_{\mu\nu} - {n_{\mu}p_{\nu}+n_{\nu}p_{\mu}\over
(n\cdot p)} + {n^{2}p_{\mu}p_{\nu}\over (n\cdot p)^{2}} +
im\epsilon_{\mu\nu\lambda} {n^{\lambda}\over (n\cdot
p)}\right]\label{7} 
\end{equation}
which can be trivially checked to be transverse to $n^{\mu}$, namely,
\begin{equation}
n^{\mu}D^{(0) ab}_{\mu\nu}(p) = 0 =
D^{(0) ab}_{\mu\nu}(p)n^{\nu}\label{8} 
\end{equation}
This observation is quite significant as we will see shortly.

\begin{figure}[t!]
\begin{center}    \epsfbox{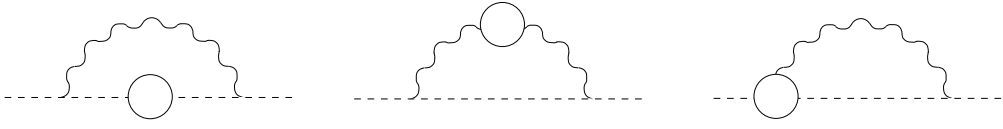}   \end{center}
\caption{Diagrams, which can contribute to the ghost self-energy,
  vanish because of Eq. (24).}
\label{fig2}
\end{figure}

Let us note that the theory described by Eq. (\ref{6}) is also invariant
under the BRST transformations of Eq. (\ref{4}). Thus, one can derive, as
usual (by adding sources as in Eq. (\ref{3a}) except for the last source),
the BRST identities for the theory, which are derived from the
master identity
\begin{equation}
\int d^{3}x\,\left({\delta\Gamma\over \delta A_{\mu}^{a}(x)}
{\delta\Gamma\over \delta K^{\mu,a}(x)} - {\delta\Gamma\over \delta
c^{a}(x)} {\delta\Gamma\over \delta L^{a}(x)} + F^{a}(x)
{\delta\Gamma\over \delta \overline{c}^{a}(x)}\right) = 0\label{9}
\end{equation}
The master identity is the same as in any other gauge. However, the
constraints following from them, in the axial gauge, are much simpler
than, say in a covariant gauge. For example, looking at the structure
of the ghost Lagrangian in Eq. (\ref{5}), we note that, in the axial gauge,
the vertex describing the coupling of the ghosts to the gluons is
proportional to $n^{\mu}$. Combined with Eq. (\ref{8}), this, then, implies
that, in the axial gauge, the ghost two point function does not receive any
quantum correction. As a result, in this gauge, the ghost wave
function renormalization is trivial,
\begin{equation}
\widetilde{Z}_{3} = 1\label{10}
\end{equation}
Similarly, it also follows that, in this gauge, the ghost-gluon
interaction vertex is not renormalized, leading to
\begin{equation}
\widetilde{Z}_{1} = 1\label{11}
\end{equation}
As a result, the standard relation following from the master identity
in Eq. (\ref{9}), in a non-Abelian gauge theory, takes the simple form
\begin{eqnarray}
{Z_{1}\over Z_{3}} & = & {\widetilde{Z}_{1}\over
\widetilde{Z}_{3}}\nonumber\\
\noalign{\kern 4pt}%
{\rm or,}\quad Z_{1} & = & Z_{3}\label{12}
\end{eqnarray}
Here, we have denoted the wave function and the vertex
renormalizations for the gauge field by $Z_{3}$ and $Z_{1}$
respectively. This relation is reminiscent of the Ward identity in an
Abelian theory. Thus, in the axial gauge, the Ward identities are
simpler, much like in the Abelian theory. However, the non-Abelian
interactions still make the structure of any amplitude much more
complex and rich.

\begin{figure}[t!]
\begin{center}    \epsfbox{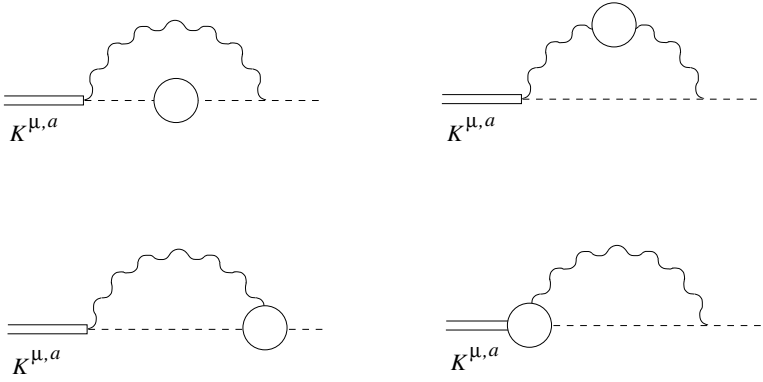} \end{center}
\caption{Diagrams, which can lead to the renormalization of the 
$K^{\mu ,\,a}\,c^a$ vertex, vanish because of Eq. (24).}
\label{fig3}
\end{figure}

Just as we see that the  ghost wave function as well as the ghost vertex
renormalizations are trivial in the axial gauge, it is equally
straightforward to show that the source terms with composite variations are
not renormalized in the axial gauge either (namely, vertices involving the
sources $K^{\mu,a}$ and $L^{a}$ receive no quantum correction). As a
result, the Ward identities following from the master identity in
Eq. (\ref{9}) take a much simpler form in the axial gauge. For example, it
follows from Eq. (\ref{9}) that, for $n\geq 2$,
\begin{eqnarray*}
 & & {\delta^{n}\Gamma\over \delta A_{\mu_{1}}^{a_{1}}(p_{1})\cdots \delta
A_{\mu_{n-1}}^{a_{n-1}}(p_{n-1})\delta A_{\mu_{n}}^{a}(p_{n})}
{\delta^{2}\Gamma\over \delta K^{\mu_{n},a}(-p_{n})\delta
c^{a_{n}}(p_{n})}\\
\noalign{\kern 4pt}%
 & = & -
\sum_{i=1}^{n-1} \int d^{3}p {\delta^{n-1}\Gamma\over \delta
A_{\mu_{1}}^{a_{1}}(p_{1})\cdots \delta
A_{\mu_{i-1}}^{a_{i-1}}(p_{i-1})\delta
A_{\mu_{i+1}}^{a_{i+1}}(p_{i+1})\cdots \delta A_{\mu_{n}}^{a}(p)}\\
\noalign{\kern 4pt}%
 &   & \qquad\qquad\quad\times{\delta^{3}\Gamma\over \delta K^{\mu_{n},a}(-p)\delta
A_{\mu_{i}}^{a_{i}}(p_{i})\delta c^{a_{n}}(p_{n})}
\end{eqnarray*}
Recalling that the vertices with the sources $K^{\mu,a}$ do not get any
quantum correction, this identity can also be rewritten in the simple
form
\begin{eqnarray}
&    p_{n,\mu_{n}}\Gamma_{a_{1}\cdots
 a_{n}}^{\mu_{1}\cdots\mu_{n}}(p_{1},\cdots, p_{n}) = \nonumber\\
%\noalign{\kern 4pt}%
&   \!\! ig\!\sum_{i=1}^{n-1} f^{aa_{i}a_{n}}
\Gamma^{\mu_{1}\cdots\mu_{n-1}}_{a_{1}\cdots a_{i-1}aa_{i+1}\cdots
a_{n-1}} (p_{1},\cdots ,p_{i-1},p_{i}+p_{n},p_{i+1},\cdots ,p_{n-1})\label{13}
\end{eqnarray}
This is clearly a much simpler identity, relating successive vertex
functions, than the identities one obtains, for example, in a covariant
gauge. Furthermore, let us note that taking the derivative with
respect to $p_{n}$ 
and setting $p_{n}=0$ in Eq. (\ref{13}), we obtain 
\begin{eqnarray}
& \Gamma^{\mu_{1}\cdots\mu_{n}}_{a_{1}\cdots a_{n}}(p_{1},\cdots
,p_{n-1},0) = \\ \nonumber
& ig \sum_{i=1}^{n-1} f^{aa_{i}a_{n}} {\partial\over
\partial p_{i,\mu_{n}}}\Gamma^{\mu_{1}\cdots \mu_{n-1}}_{a_{1}\cdots
a_{i-1}aa_{i+1}\cdots a_{n-1}}(p_{1},\cdots ,p_{n-1})\label{14}
\end{eqnarray}

The two point function in the full theory, in a generalized axial
gauge ($\xi\neq 0$), can be parameterized, consistent with the BRST
identities, as
\begin{eqnarray}
\Pi^{\mu\nu, ab}(p) & = &
\delta^{ab}\left[-\left(\eta^{\mu\nu}-p^{\mu}p^{\nu}\right)
(1+\Pi_{1}(p)) + im\epsilon^{\mu\nu\lambda}p_{\lambda}
(1+\Pi_{2}(p))\right.\nonumber\\
\noalign{\kern 4pt}%
 &  & \qquad \left. + (p^{\mu} - {p^{2}n^{\mu}\over (n\cdot p)})
(p^{\nu} - {p^{2}n^{\nu}\over (n\cdot p)}) \Pi_{3}(p) - {1\over \xi}
n^{\mu}n^{\nu}\right] 
\end{eqnarray}
The self-energy (which, by definition, is the two point function without
the tree level terms) is clearly transverse with respect to momentum.
Let us note that relation (\ref{14}) must hold for both parity
conserving  as
well as parity violating parts of the amplitudes separately. Thus,
looking at the parity violating part of the three point amplitude, we
obtain, (Note that we have identified $\Gamma^{\mu\nu,ab} =
\Pi^{\mu\nu,ab}$.)
\begin{eqnarray}
\epsilon_{\mu\nu\lambda}\Gamma^{\mu\nu\lambda}_{abc} & = & ig
f^{dbc}\epsilon_{\mu\nu\lambda}\left. {\partial\over \partial
p_{\lambda}}\Pi^{\mu\nu,ad}(p)\right|_{p=0}\nonumber\\
\noalign{\kern 4pt}%
{\rm or,}\quad f^{abc} (1 + \Pi_{2}(0)) & = & {1\over 6im}
f^{dbc}\epsilon_{\mu\nu\lambda}\left.{\partial\over \partial
p_{\lambda}}\Pi^{\mu\nu,da}(p)\right|_{p=0}\nonumber\\
\noalign{\kern 4pt}%
 & = & {1\over 6mg} \epsilon_{\mu\nu\lambda}
 \Gamma^{\mu\nu\lambda}_{abc}(0,0,0)\label{15}
\end{eqnarray} 
This relation is quite crucial in that it relates the Chern-Simons
coefficient in the axial gauge, at any order, to the parity violating
part of  the three
gluon vertex (with vanishing momenta) at the same order. Thus, one can
give a diagrammatic representation for the Chern-Simons coefficient in
the axial gauge, which is very convenient for studying an all
order proof of the generalization of the Coleman-Hill result to
non-Abelian theories. It is also clear from this identification that
the choice of an infrared safe gauge is crucial because the
Chern-Simons coefficient is related to the three gluon amplitude with
all external momenta vanishing. 

In a non-Abelian Chern-Simons theory, as we have argued earlier, the
ratio ${4\pi m\over g^{2}}$ represents a physical quantity. This is
known to be true from the following facts, namely, in the leading order in
${1\over m}$ expansion, i) it is this ratio which determines the
dimensionality of the Chern-Simons Hilbert space {\cite{15}}
and  ii) this ratio is related to the coefficient 
of the WZWN action which represents the central charge of the 
corresponding current algebra {\cite{16}}. It is also this
ratio (see Eq. (\ref{3})) which needs to be quantized for {\it large
gauge} invariance of the theory. In the
full quantum theory, however, this ratio changes as
\begin{equation}
{4\pi m\over g^{2}} \rightarrow \left({4\pi m\over g^{2}}\right)_{\rm
ren} = Z_{m}\left({Z_{3}\over Z_{1}}\right)^2\left({4\pi m\over
g^{2}}\right)\label{16}
\end{equation}
where $Z_{3}$ and $Z_{1}$ are the wave function and the vertex
renormalization constants for the gauge field (as we have defined
earlier), while $Z_{m}$ represents the renormalization of the
Chern-Simons coefficient. By definition, of course,
$Z_{m}=(1+\Pi_{2}(0))$ and since in the axial gauge we have
$Z_{1}=Z_{3}$ (see Eq. (\ref{12})), it follows that
\begin{equation}
\left({4\pi m\over g^{2}}\right)_{\rm ren} = Z_{m} \left({4\pi m\over
g^{2}}\right) = (1+\Pi_{2}(0))\left({4\pi m\over g^{2}}\right)\label{17}
\end{equation}
Since this is a physical quantity, it follows that, in the axial
gauge, the induced Chern-Simons coefficient takes on a physical
meaning. 

In fact, let us next show that this expectation is indeed true and
that the Chern-Simons coefficient is independent of $n^{\mu}$ in the
axial gauge. To prove this, let us add to our Lagrangian density the
following source terms.
\begin{eqnarray}
{\cal L}_{source} & = & J^{\mu,a}A_{\mu}^{a} + J^{a}F^{a} + i
(\overline{\eta}^{a}c^{a} - \overline{c}^{a} \eta^{a})
 + K^{\mu,a}D_{\mu}c^{a}\nonumber\\
  &  & \qquad + L^{a} (-{1\over 2}f^{abc}c^{b}c^{c}) +
H^{\mu}\overline{c}^{a} A_{\mu}^{a}
\end{eqnarray}
Thus, defining
\[
{\cal L}_{total} = {\cal L}  + {\cal L}_{source} 
\]
we note that, under a BRST transformation (see Eq. (\ref{4})),
\begin{equation}
\delta {\cal L}_{total} = \omega\left[J^{\mu,a}(D_{\mu}c^{a}) -
i\overline{\eta}^{a}(-{1\over 2}f^{abc}c^{b}c^{c}) + iF^{a}\eta^{a} +
H^{\mu}{\partial{\cal L}_{total}\over \partial n^{\mu}}\right]
\end{equation}
Thus, as before, making a field redefinition inside the path integral,
which coincides with a BRST transformation, we can derive the equation
which describes how the effective action changes with $n^{\mu}$. Let
us simply note the result here,
\begin{eqnarray}
{\partial\Gamma\over \partial n^{\mu}} & = & -\int
d^{3}p\left[{\delta\Gamma\over \delta
A_{\nu}^{a}(-p)}{\delta^{2}\Gamma\over \delta K^{\nu,a}(p)\delta
H^{\mu}(-p)}+{\delta F^{a}(p)\over \delta H^{\mu}(p)}{\delta\Gamma\over
\delta \overline{c}^{a}(-p)}\right.\nonumber\\
\noalign{\kern 4pt}%
 &  & \qquad\qquad \left. - {\delta\Gamma\over \delta
c^{a}(-p)}{\delta^{2}\Gamma\over \delta L^{a}(p)\delta
H^{\mu}(-p)}\right]
\end{eqnarray}
This is the master identity from which we obtain,
\begin{eqnarray}
 {\partial\over \partial
n^{\rho}}\left(\epsilon_{\mu\nu\lambda}{\delta^{3}\Gamma\over \delta
A_{\mu}^{a}\delta A_{\nu}^{b}\delta
A_{\lambda}^{c}}\right)_{p_{1},p_{2},p_{3}=0}  =  
-\epsilon_{\mu\nu\lambda}\left[{\delta^{4}\Gamma\over \delta 
A_{\mu}^{a}\delta A_{\nu}^{b}\delta A_{\lambda}^{c}\delta
A_{\sigma}^{d}}{\delta^{2}\Gamma\over \delta K^{\sigma,d}\delta
H^{\rho}}\right.\nonumber\\
% \noalign{\kern 4pt}%
     + \left. {\delta^{3}\Gamma\over \delta
 A_{\mu}^{a}\delta A_{\nu}^{b}\delta
 A_{\sigma}^{d}}{\delta^{3}\Gamma\over \delta K^{\sigma,d}\delta
 H^{\rho}\delta A_{\lambda}^{c}}\right]_{p_{1},p_{2},p_{3}=0}
%\nonumber\\
%\noalign{\kern 4pt}%
 +  {\rm permutations}\label{18}
\end{eqnarray}
where the restriction on the right hand side stands for setting all
the field variables as well as momenta equal to zero.

\begin{figure}[ht!]
\begin{center}    \epsfbox{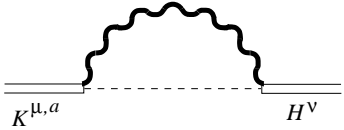} \end{center}
\caption{Diagram representing the possible mixing of the sources 
$K^{\mu ,\,a}$ and $H^\nu$. The bold wavy line represents the complete
  gauge field propagator. The diagram vanishes because of color identities.}
\label{fig4}
\end{figure}

It is easy to see that, to all orders in the quantum theory, we cannot
have a vertex of the form $K^{\mu,a}H^{\nu}$. In fact, let us 
recall that in the axial gauge the ghost propagator as well as the
ghost vertex do not renormalize. Similarly, just as we noted that the
vertices involving the source $K^{\mu,a}$ do not renormalize, we can
also show that the vertex involving $H^{\mu}$ does not renormalize in
the quantum theory either. It follows from this that the only diagram that
can give rise to a mixing of the sources $K^{\mu,a}$ and $H^{\nu}$ is
as shown in Fig. 4. From the fact
that the vertex $K^{\mu,a}A_{\nu}^{b}c^{c}$ is anti-symmetric in the
internal indices while the vertex $H^{\mu}A_{\nu}^{a}c^{b}$ and the
gauge propagator are symmetric in the internal symmetry indices, it
follows  that
this diagram vanishes. (Alternately, such a vertex, if it existed,
would involve a single internal index, which is impossible to
construct from the structures present in the theory.)

\begin{figure}[ht!]
\begin{center}    \epsfbox{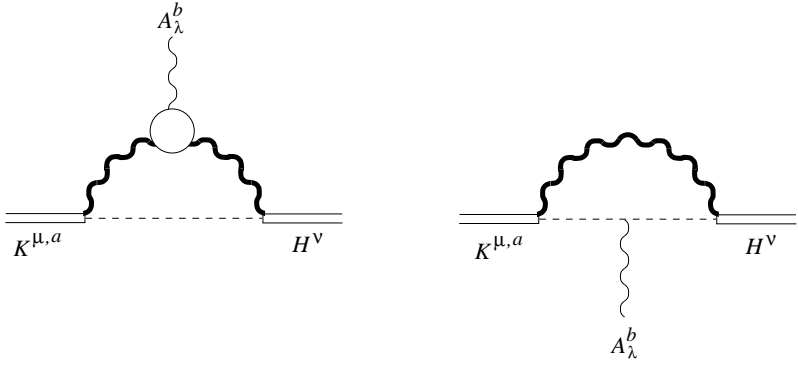} \end{center}
\caption{Diagrams contributing to a vertex of the type 
$K_\mu^a\, H^\nu\, A_\lambda^b$. The sum of the two diagrams vanishes.}
\label{fig5}
\end{figure}

Let us next analyze if a vertex of the form
$K^{\mu,a}H^{\nu}A_{\lambda}^{b}$ can be generated in the quantum
theory. To that extent, let us note the following simple identity in
the axial gauge.
\begin{equation}
{\partial\over \partial p_{\mu}}\widetilde{D}(-p) = 
{i\over
g}
\widetilde{D}(-p)\widetilde{\Gamma}^{\mu}(-p,0,p)\widetilde{D}(-p)\label{19}  
\end{equation}
where we have represented the ghost propagator by $\widetilde{D}$ and
the ghost vertex by $\widetilde{\Gamma}$ without the internal symmetry
indices (remember that these do not
receive any quantum correction and, therefore, coincide with their
tree level forms.), namely,
\begin{eqnarray*}
i\widetilde{D}^{ab}(p) & = & i\delta^{ab}\widetilde{D}(p) =
 {\delta^{ab}\over (n\cdot p)}\\
\noalign{\kern 4pt}%
\widetilde{\Gamma}^{\mu,abc} (-p,0,p) & = &
  f^{abc}\widetilde{\Gamma}^{\mu}(-p,0,p) = g f^{abc} n^{\mu}
\end{eqnarray*}
The identity, in Eq. (\ref{19}), is reminiscent
of the Abelian identity involving fermion lines, namely, it says that
differentiating the ghost propagator is equivalent to introducing a
photon line with zero momentum. Using these, as well as the fact that
the vertices involving a $K^{\mu,a}$ or a $H^{\mu}$ do not
renormalize, we note that the only diagrams which can generate a
vertex of the type $K^{\mu,a}H^{\nu}A_{\lambda}^{b}$ are as shown in
Fig. 5. Evaluating these at zero external momenta, we obtain
\begin{eqnarray}
  & \sim & \int
 d^{3}p\left[f^{aa'a''}\widetilde{D}(-p)D_{\mu\mu'}^{a'b'}(p)\Gamma^{\mu'\nu'
\lambda}_{b'b''b}(p,-p,0)D_{\nu'\nu}^{b''a''}(p)\right.\nonumber\\
\noalign{\kern 4pt}%
 &  & \qquad\left. +
 f^{aa'a''}f^{a''bb'}\widetilde{D}(-p)\widetilde{\Gamma}^
{\lambda} (-p,0,p)\widetilde{D}(-p) D_{\mu\nu}^{a'b'}(p) \right]\label{19a}
\end{eqnarray}
From Eq. (\ref{14}), we note that we can write
\begin{equation}
D_{\mu\mu'}^{aa'}(p)\Gamma^{\mu'\nu'\lambda}_{a'c'b}(p,-p,0)D_{\nu'\nu}^{c'c}
(p) = ig f^{dbc} {\partial\over \partial
p_{\lambda}}D_{\mu\nu}^{da}(p)\label{20}
\end{equation}
Using this, as well as Eq. (\ref{19}), the contribution of Fig. 5 in
Eq. (\ref{19a})  can be simplified as
\begin{eqnarray}
& \sim & \int d^{3}p \left[f^{aa'a''}\widetilde{D}(-p)(ig
f^{dba''}){\partial\over \partial
p_{\lambda}}D_{\mu\nu}^{da'}(p)\right.\nonumber\\
\noalign{\kern 4pt}%
 &  & \qquad \left. + f^{aa'a''}f^{a''bb'}(-ig)({\partial\over \partial
 p_{\lambda}} \widetilde{D}(-p))
 D_{\mu\nu}^{a'b'}(p)\right]\nonumber\\
\noalign{\kern 4pt}%
 & = & ig f^{aa'a''}f^{dba''} \int d^{3}p {\partial\over \partial
 p_{\lambda}}(\widetilde{D}(-p)D_{\mu\nu}^{da'}(p))\nonumber\\
\noalign{\kern 4pt}%
 & = & 0
\end{eqnarray}
In other words, a vertex of the kind $K^{\mu,a}H^{\nu}A_{\lambda}^{b}$
is not generated in the full quantum theory. It follows now that, in
such a case, Eq. (\ref{18}) leads to
\begin{equation}
{\partial\over \partial
n^{\rho}}\left(\epsilon_{\mu\nu\lambda}{\delta^{3}\Gamma\over \delta
A_{\mu}^{a}\delta A_{\nu}^{b}\delta
A_{\lambda}^{c}}\right)_{p_{1},p_{2},p_{3}=0} = 0\label{21} 
\end{equation}
Namely, the Chern-Simons coefficient is independent of the choice of
$n^{\mu}$ as was stated. This is consistent with our observation that
the Chern-Simons coefficient takes on a physical meaning in the axial
gauge. 

\section{One-loop calculation:}

In this section, let us check explicitly, at the one loop level, that
the Chern-Simons coefficient is independent of $n^{\mu}$ as was shown
from general arguments, in the previous section. Let us recall that
the Chern-Simons coefficient, in the axial gauge, can be related to
the parity violating part of the three gluon amplitude with all
external momenta vanishing. At one loop level, there are two such
diagrams  that
would contribute to the Chern-Simons coefficient -- one is the triangle
graph and the other involving the quartic interaction vertex as shown
in Figs. 6a and 6b. Let us
first look at the simpler of the two graphs, namely, the one involving
the quartic interaction vertex. The contribution coming from this
diagram (contracted with $\epsilon_{\mu\nu\lambda}$) can be written as
\begin{eqnarray}
 I_{(6b)}^{abc} =   %\nonumber\\
%\noalign{\kern 4pt}%
% \!\!\! 
\epsilon_{\mu\nu\lambda} \int {d^{3}p\over
 (2\pi)^{3}}\,D_{\nu'\mu'}^{(0)b'a'}(p)
 \Gamma^{(0)\mu'\mu''\mu}_{a'a''a}(p,-p,0) \nonumber \\
 \times D_{\mu''\lambda'}^{(0)a''c'}(p)
 \Gamma^{(0)\lambda'\nu'\nu\lambda}_{c'b'bc}(p,-p,0,0)  \nonumber\\ 
%\noalign{\kern 4pt}%
  = ig \epsilon_{\mu\nu\lambda} \int {d^{3}p\over
 (2\pi)^{3}}\,f^{dac'}({\partial\over \partial
 p_{\mu}}D_{\nu'\lambda'}^{db'}(p))
 \Gamma^{(0)\lambda'\nu'\nu\lambda}_{c'b'bc}(p,-p,0,0) \nonumber\\
%\noalign{\kern 4pt}%
  = ig \int {d^{3}p\over (2\pi)^{3}} {\partial\over \partial
 p_{\mu}}\left(f^{dac'}
 D_{\nu'\lambda'}^{db'}(p)
\Gamma^{(0)\lambda'\nu'\nu\lambda}_{c'b'bc}(p,-p,0,0)\right)
  = 0  \label{22}
\end{eqnarray}
Here, we have used Eq. (\ref{20}) as well as the fact the the tree level four
point  vertex is
independent of momenta and hence can be taken inside the
differentiation. This shows that, of the two diagrams that can
possibly contribute to the Chern-Simons term at one loop, the one with
the quartic interaction vertex vanishes. As we will see later, this
property generalizes in a simple manner to higher loops. 

\begin{figure}[t!]
    \epsfbox{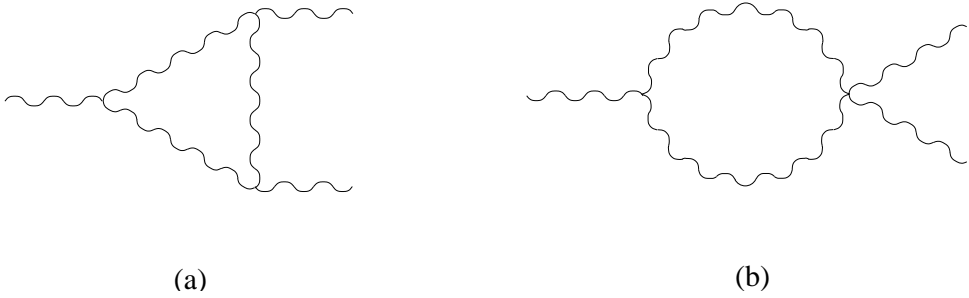}
\caption{Diagrams which can contribute to the Chern-Simons coefficient at
  one loop.}
\label{fig6}
\end{figure}

This analysis shows that the entire contribution, at one-loop level,
to the Chern-Simons coefficient would come from the triangle
diagram in Fig. 6a. The triangle diagram can be simplified slightly by
the use of the
identity (\ref{20}), but the evaluation is tedious and leads to the
contribution (when contracted with $\epsilon_{\mu\nu\lambda}$)
\begin{eqnarray}
 I_{(6a)}^{abc} % \nonumber\\ 
& = &\!\!\! - Ng^{3} f^{abc} \int {d^{3}p\over (2\pi)^{3}}\,\left[{12im\over
(p^{2}-m^{2})^{2}}  \right. \nonumber\\ 
&+& {\partial\over \partial
p_{\mu}}\left({8im^{3}p_{\mu}\over (p^{2}-m^{2})^{3}} +
{5imn_{\mu}\over (p^{2}-m^{2})(n\cdot p)} \right.\nonumber\\
%\noalign{\kern 4pt}%
 &  & 
%\qquad\qquad\qquad 
\left.\left. - {4im^{3}n_{\mu}\over
 (p^{2}-m^{2})^{2}(n\cdot p)} - {8im^{5}n_{\mu}\over
 (p^{2}-m^{2})^{3}(n\cdot p)}\right)\right]\nonumber\\
\noalign{\kern 4pt}%
 & = & - Ng^{3} f^{abc} \int {d^{3}p\over (2\pi)^{3}}\,{12im\over
 (p^{2}-m^{2})^{2}}\nonumber\\
\noalign{\kern 4pt}%
 & = & - Ng^{3} f^{abc} \left(-{6\over 4\pi}\right) = 6g\, {g^{2}N\over
 4\pi}\,f^{abc}
\end{eqnarray}
It follows now from Eq. (\ref{15}) that, at one-loop,
\begin{equation}
\Pi_{2}^{(1)}(0) = {g^{2}N\over 4\pi m}
\end{equation}
Alternately, the shift in the tree level Chern-Simons coefficient, due
to one-loop effects, is
\begin{equation}
m\Pi_{2}^{(1)}(0) = {g^{2}\over 4\pi} N
\end{equation}
There are several things to note from this calculation. First, we see
explicitly that the one-loop Chern-Simons coefficient is independent
of $n^{\mu}$, consistent with the proof of the earlier section. Second,
since the wave function and the vertex renormalizations for the gauge
field are identical in the axial gauge (see Eqs. (\ref{12}) and
(\ref{17})), in this gauge, at one loop,
\begin{equation}
\left({4\pi m\over g^{2}}\right)^{(1)}_{\rm ren} = {4\pi m\over g^{2}}
+ N\label{23}
\end{equation}
In other words, this ratio shifts by $N$ (of $SU(N)$) at one
loop. This is exactly what was also found from a calculation in the
covariant Landau gauge {\cite{7}}, 
which re-confirms that this is indeed a gauge independent quantity.
From an algebraic point of view, one can give  a meaning to
the one-loop  shift of the Chern-Simons coefficient as the product of
the spin with the dual Coxeter number of the group
{\cite{16,17}}.

Let us note here that, in general, if we choose a general gauge fixing
of the kind
\begin{equation}
{\cal L}_{gf} = - {1\over 2} (\Lambda^{\mu}A_{\mu}^{a})^{2}
\end{equation}
then, the tree level propagator for the gauge field, in this gauge, can be
determined to be ($\Lambda^{\mu}$ is assumed to be independent of the
gauge field.)
\begin{eqnarray}
D_{\mu\nu}^{(0) ab}(p)\!\! & = &\!\!  {\delta^{ab}\over
p^{2}-m^{2}}\!\left[\eta_{\mu\nu} - {p_{\mu}\Lambda_{\nu}(p)\over
(p\cdot \Lambda(p))} - {p_{\nu}\Lambda_{\mu}(-p)\over (p\cdot
\Lambda(-p))} + \right.\nonumber\\
&  & \qquad\left. +
{p_{\mu}p_{\nu}\Lambda(p)\cdot \Lambda(-p)\over
(p\cdot \Lambda(p))(p\cdot \Lambda(-p))} %\right.\nonumber\\
%% \noalign{\kern 4pt}%
%% &  & \qquad\left. +
+ im\epsilon_{\mu\nu\lambda}\,{\Lambda^{\lambda}(-p)\over (p\cdot
 \Lambda(-p))}\right]  
\nonumber\\
&  & \qquad +
{\delta^{ab}p_{\mu}p_{\nu}\over (p\cdot
 \Lambda(p)) (p\cdot \Lambda(-p))}\label{24}
\end{eqnarray}
The covariant gauge propagator would follow from this with the choice
\[
\Lambda^{\mu}(p) = -{ip^{\mu}\over \sqrt{\xi}}
\]
while the propagator in the general axial gauge would follow from the
choice
\[
\Lambda^{\mu}(p) = {n^{\mu}\over \sqrt{\xi}}
\]
But, in fact, we can have more interesting gauge choices with
\begin{equation}
\Lambda^{\mu}(p) = - {1\over \sqrt{\xi}} (i\beta p^{\mu} +
(1-\beta)n^{\mu})
\end{equation}
Here $\beta$ is an arbitrary parameter and we note that such a choice
of gauge allows us to interpolate between the covariant and the axial
gauges. Namely, when $\beta =1$, we have the covariant gauge, whereas
for $\beta =0$, we have the general axial gauge. The tree level propagator, in
this interpolating gauge takes the form
\begin{eqnarray}
D_{\mu\nu}^{(0) ab}(p) & = &  {\delta^{ab}\over
p^{2}-m^{2}}\left[\eta_{\mu\nu} - {p_{\mu}(i\beta
p_{\nu}+(1-\beta)n_{\nu})\over (i\beta p^{2}+(1-\beta)(n\cdot
p))}\right.\nonumber\\
\noalign{\kern 4pt}%
 &  & \qquad - {p_{\nu}(i\beta p_{\mu}-(1-\beta)n_{\nu})\over (i\beta
 p^{2} -(1-\beta)(n\cdot p))}\nonumber\\
\noalign{\kern 4pt}%
 &  & \qquad + {p_{\mu}p_{\nu}(i\beta p+(1-\beta)n)\cdot (i\beta
 p-(1-\beta)n)\over (i\beta p^{2}+(1-\beta)(n\cdot p))(i\beta
 p^{2}-(1-\beta)(n\cdot p))}\nonumber\\
\noalign{\kern 4pt}%
 &  & \qquad\left. + im\epsilon_{\mu\nu\lambda}\,{(i\beta
 p^{\lambda}-(1-\beta)n^{\lambda})\over (i\beta
 p^{2}-(1-\beta))(n\cdot p))}\right]\nonumber\\
\noalign{\kern 4pt}%
 &  & - {\delta^{ab}\xi p_{\mu}p_{\nu}\over (i\beta
 p^{2}+(1-\beta)(n\cdot p))(i\beta p^{2}-(1-\beta)(n\cdot p))}
\end{eqnarray}
For $\xi=0$, this provides an infrared safe gauge, which interpolates
between the Landau gauge and the axial gauge. We note that, following
earlier discussions, we can write an identity which will describe the
$\beta$ dependence of various amplitudes. Thus, adding a source
Lagrangian density of the form,
\begin{eqnarray*}
{\cal L}_{source} & = & J^{\mu,a}A_{\mu}^{a} + J^{a}F^{a} +
i(\overline{\eta}^{a} c^{a}-\overline{c}^{a}\eta^{a}) +
K^{\mu,a}(D_{\mu}c^{a}) \\
\noalign{\kern 4pt}%%%%crr 
 &  & \qquad + L^{a}(-{1\over 2}f^{abc}c^{b}c^{c}) +
 H(\partial^{\mu}\overline{c}^{a}-n^{\mu}\overline{c}^{a}) A_{\mu}^{a}
\end{eqnarray*}
we can derive the master identity describing the $\beta$ dependence of
the effective action to be of the form,
\begin{eqnarray}
{\partial\Gamma\over \partial\beta} & = & - \int d^{3}x
d^{3}y\left({\delta\Gamma\over \delta
A_{\mu}^{a}(x)}{\delta^{2}\Gamma\over \delta K^{\mu,a}(x)\delta H(y)}
+{\delta\Gamma\over \delta c^{a}(x)}{\delta^{2}\Gamma\over \delta
L^{a}(x)\delta H(y)}\right.\nonumber\\
\noalign{\kern 4pt}%
 &  & \qquad\qquad\qquad \left. - {\delta F^{a}(x)\over \delta
H(y)}{\delta\Gamma\over \delta\overline{c}^{a}(x)}\right)
\end{eqnarray}
While we have not done this, we believe that it is possible to show
from this that the ratio ${4\pi m\over g^{2}}$ is independent of
$\beta$, as has been explicitly seen from the one loop
calculation.

\section{Proof of the main result:}

In this section, we will argue that, in the axial gauge, the
Chern-Simons coefficient receives no contribution beyond one loop,
when the {\it small gauge} Ward identities hold and the amplitudes are
analytic in the momentum variables. This, therefore, would be the
generalization
of the Coleman-Hill result to non-Abelian gauge theories. It will
follow from this result that the ratio ${4\pi m\over g^{2}}$ has no quantum
correction beyond one loop in any gauge.

To simplify our proof, let us employ a compact notation where we treat
the amplitudes as matrices in the Lorentz and internal symmetry
space. Thus, we define $\Pi$, $D$, $\Gamma^{\lambda}$ and
$\Gamma^{\nu\lambda}$ to represent respectively the complete two point
function, the propagator, the three point and the four point vertex
functions for the gauge fields. In this notation, then, we have
\begin{equation}
\Pi D = -1\label{25}
\end{equation}
and, furthermore, when the momentum associated with the free index
vanishes, we can obtain, using this, from Eq. (\ref{14})
\begin{eqnarray}
\Gamma^{\lambda} & = & ig \partial^{\lambda}\Pi\nonumber\\
\noalign{\kern 4pt}%
{\rm or,}\quad D\Gamma^{\lambda}D & = & ig D(\partial^{\lambda}\Pi)D =
ig \partial^{\lambda}D\label{26}
\end{eqnarray}
Here and in what follows, $\partial^{\lambda}$ represents the
derivative with respect to the appropriate momentum and we have
ignored writing out the explicit internal indices for
simplicity. (Namely, the internal symmetry factors simply come out of
the integrals and are not relevant to our proof as will become evident
shortly.) We recognize Eq. (\ref{26}) as the relation in
Eq. (\ref{20}) in our compact notation.

\begin{figure}[ht!]
    \epsfbox{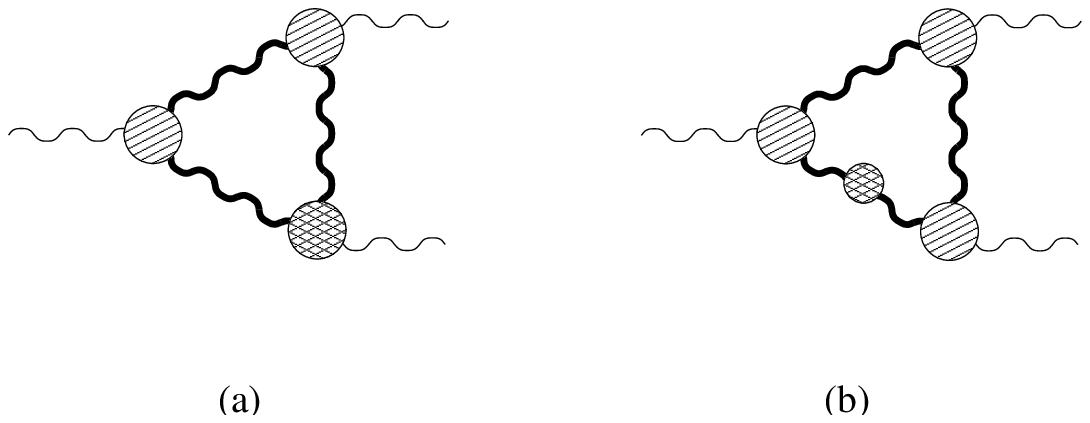}
\caption{A class of diagrams that can contribute to higher order
corrections of the Chern-Simons coefficient in the axial gauge.  
The hatched vertices and the bold internal lines represent
respectively the three point vertices and the propagators which
include all the corrections up to $n$-loop order. The cross-hatched 
vertex and the  cross-hatched blob (self-energy) in the internal 
propagator, include all the corrections up to $(n+1)$-loop order.}
\label{fig7}
\bigskip 
\bigskip
\bigskip 
\bigskip
%\end{figure}
%
%\begin{figure}[h!]
\begin{center}
    \epsfbox{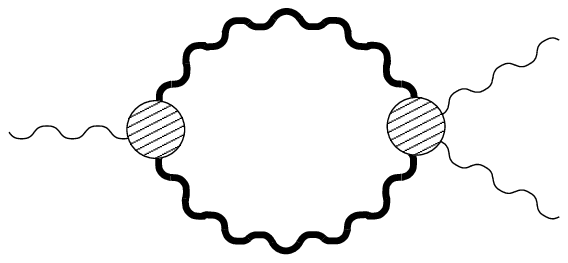}
\end{center}
\label{fig8}
\caption{Second class of diagrams that can contribute to higher order
corrections of the Chern-Simons coefficient in the axial gauge. Here,
the diagram involves vertices and propagators of the full theory.}
\end{figure}
 
In analyzing the higher loop contributions to the Chern-Simons
coefficient, let us note that there are two possible classes of
diagrams which are of interest and are shown in Figs. 7 and 8. Let us
look at the class of diagrams in Figs. 7a and 7b with all external
momenta vanishing. Here the hatched vertices and the bold internal
lines represent respectively the three point vertices and the
propagators, which include all the corrections up to $n$-loop order,
with $n=0,1,2,\cdots$. The cross-hatched vertex, on the other hand,
includes all the corrections up to $(n+1)$-loop order, starting from
one-loop (namely, it does not contain the tree level term). Similarly,
the cross-hatched loop in the internal propagator stands for the
self-energy, which includes all the corrections up to $(n+1)$-loops
(by definition, the self-energy does not contain the tree level two
point function). We will put an overline, on these two factors, just
to emphasize that they do not contain the tree level contribution. It
is clear now that, by construction, the diagrams in Figs. 7a and 7b
lead to contributions only at two loops and higher. Furthermore, from
the definition given above, we can write, in the notation described
earlier
\begin{equation}
\overline{\Gamma}^{\lambda} = ig \partial^{\lambda}\overline{\Pi}
\label{27}
\end{equation}

The contributions from the diagrams in Figs. 7a and 7b, when
contracted with $\epsilon_{\mu\nu\lambda}$, would yield a part of the
$(n+2)$-loop corrections to the Chern-Simons term and takes the form
\begin{eqnarray}
I^{(n+2)}_{(7a)+(7b)}  =  
{\rm Tr} \int d^{3}q\,\epsilon_{\mu\nu\lambda}
\left[\left(D \Gamma^{\mu} D \Gamma^{\nu}  D
\left(
\overline{\Gamma}^{\lambda}+\Gamma^{\lambda} D\overline{\Pi}
\right)\right)^{(n+1)}
+ {\rm cyclic} \right]\nonumber \\
  =  - i\,g^{3}\,{\rm Tr} \int d^{3}q\,\epsilon_{\mu\nu\lambda}
\left[\left(\partial^\mu D \partial^\nu \Pi
\left(
D\partial^\lambda \overline{\Pi} +
\partial^\lambda D\overline{\Pi}\right)
\right)^{(n+1)} + {\rm cyclic}\right]\nonumber \\
  =   -i\,g^{3}\,{\rm Tr} \int d^{3}q\,\epsilon_{\mu\nu\lambda}
\left[\partial^\lambda
\left(
\partial^\mu D\partial^\nu\Pi D\overline{\Pi}
\right)^{(n+1)} + {\rm cyclic}\right]= 0\label{28} 
\end{eqnarray}
for all $n=0,1,2,\cdots$, where the superscript $(n+1)$ stands for the
order of the terms in the expression. Here, \lq\lq Tr'' denotes trace
over the matrix indices in the Lorentz space and we have used the
identities in Eqs. (\ref{25})-(\ref{27}) in deriving
Eq. (\ref{28}). (There are also matrix indices associated with internal
symmetry and these are not traced, but it is clear that they are not
relevant for our argument.) We note that, because of the $\epsilon$
tensor, the factor inside the divergence in the integrand picks out
only parity violating terms in the amplitude, which converge
sufficiently rapidly to zero as $q\rightarrow \infty$. This shows that
all the higher loop contributions to the Chern-Simons coefficient,
coming from the class of diagrams in Figs. 7a and 7b, vanish.

We can, similarly, show that all the contributions, to the
Chern-Simons coefficient, coming from the class of diagrams in Fig. 8
identically vanish. We have already seen it explicitly in our one-loop
calculation in the previous section. Here, we show that it is true for
this class of diagrams at any loop. Let us note that, from the
identities in Eq. (\ref{14}), we can also express the four point
function with two external momenta vanishing, in terms of the three
point vertex with one external momentum vanishing in the compact form
as (all momenta are incoming)
\begin{equation}
\Gamma^{\nu\lambda} (q,-q;0,0) = ig \left[{\partial\over \partial
q_{\lambda}} \Gamma^{\nu}(q,-q';q-q')\right]_{q'=q} \equiv ig
\partial^{\lambda}\Gamma^{\nu}\label{29}
\end{equation}
where, again, we have suppressed the internal indices and we follow
the convention that the momenta associated with the indices
$\nu,\lambda$ of the four point vertex as well as that associated with
the index $\nu$ of the three point vertex vanish. Written out
explicitly, the right hand side of Eq. (\ref{29}) would involve two
terms with different distributions of the internal symmetry indices,
as is clear from Eq. (\ref{14}). However, as we have emphasized
earlier, the internal symmetry factors are not very relevant to the
proof of our result.

With these, let us look at the 
class of graphs in Fig. 8, with all external momenta vanishing. 
As opposed  to the diagrams in Fig. 7a and 7b, here all the 
vertices and the propagators include corrections to all orders
(namely, they are the full vertices and propagators of the theory). With the
use of Eqs. (\ref{26}) and (\ref{29}),
the contraction of $\epsilon_{\mu\nu\lambda}$ with the amplitude in
Fig.   yields
\begin{eqnarray}
I_{(8)} & = & {\rm Tr} \int d^{3}q\,\epsilon_{\mu\nu\lambda}
D\Gamma^{\mu}D\Gamma^{\nu\lambda}\nonumber
= - g^{2}\,{\rm Tr} \int d^{3}q\,\epsilon_{\mu\nu\lambda}\,
\partial^{\mu}D\partial^{\lambda}\Gamma^{\nu} \nonumber \\
& = &  -g^{2}\,{\rm Tr} \int
d^{3}q\,\partial^{\mu}\left(\epsilon_{\mu\nu\lambda}
D\partial^{\lambda}\Gamma^{\nu}\right) =  0\label{30} 
\end{eqnarray}
Once again, the integrand in Eq. (\ref{30}) is sufficiently convergent
(because it involves only the parity violating parts of the
amplitude) so that the integral vanishes.

Since these are all the diagrams that can contribute to the higher loop
corrections of the Chern-Simons coefficient, we have shown that, in a
Yang-Mills-Chern-Simons theory, the Chern-Simons coefficient, in the axial
gauge, does  not receive any correction beyond the one loop
order.  In other words,  much
like the proof in  the Abelian theory, we have used the
non-Abelian Ward  identities in the
axial gauge, together with the analyticity of the amplitudes in momentum
space, to show that the Chern-Simons coefficient has no quantum correction
beyond one-loop in this gauge. (We have explicitly checked that, in
the axial gauge, the two loop corrections of the Chern-Simons
coefficient do add up to zero.) In theories where these assumptions are
valid, we will expect our proof to hold true. On the other hand, if
either of these assumptions is violated, the proof is expected to
break down, which is quite similar to the case in the
Abelian theory. Thus, for example, in an Abelian theory with charged
massless  particles, infrared
divergences invalidate the second assumption
{\cite{18}}. Similarly, at finite
temperature, it is known that the amplitudes are non-analytic in the
energy-momentum variables {\cite{10}}
and, consequently, the Coleman-Hill result is known to be 
violated in this case {\cite{19}}.

It is worth discussing the implications of this result in some
detail. After all, we have already argued, in section {\bf 2}, that the
Chern-Simons coefficient is, in general, gauge dependent. Therefore,
even if it has no higher loop corrections in the axial gauge, this may
not hold in other gauges. For example, in  references \cite{refC,refD},
the non-abelian Chern-Simons theory was investigated by the use of
gauge invariant regularizations in covariant gauges, and was argued
that the higher order radiative corrections are finite.
Let us note here that (see Eq. (\ref{17}) and the
discussion there), since the Chern-Simons coefficient does not
receive any higher loop correction in the axial gauge, it implies
that, in this gauge, the ratio
${4\pi m\over g^{2}}$ also does not receive any higher loop
correction. On the other  hand, as we have argued,
this ratio is a
gauge independent quantity. Consequently, our result can also be
understood as saying that, in any infrared safe gauge, the ratio ${4\pi m\over
g^{2}}$ does not receive any contribution beyond one loop. This,
in fact, is the appropriate generalization of the Coleman-Hill result
to non-Abelian theories. In particular, we note from Eq. (\ref{16})
that, since in a non-axial type gauge such as the Landau gauge, 
$Z_{1}\neq Z_{3}$, in such gauges, the Chern-Simons coefficient
($Z_{m}$) will be corrected at higher loops, but in such a way that the
ratio ${4\pi m\over g^{2}}$ is unrenormalized beyond one loop.

Such a result has, of course, been expected and predicted 
{\cite{5,7}}. In fact, there is a plausibility argument 
for this, based on {\it large gauge} invariance of the theory in the 
following way. The only dimensionless
ratio, in this theory, is ${g^{2}\over 4\pi m}$ where $4\pi$ is a simple
normalization. Therefore, one can use this as a perturbation parameter
and write
\begin{equation}
\left({4\pi m\over g^{2}}\right)_{\rm ren} = {4\pi m\over
g^{2}}\,\sum_{n=0}^{\infty} a_{n}(N)\left({g^{2}\over 4\pi
m}\right)^{n}\label{31}
\end{equation}
with $a_{0}(N)=1$ and, as we have seen, $a_{1}(N)=N$. On the other
hand, the invariance of the Chern-Simons theory under {\it large
gauge} transformations requires that this ratio be quantized (see
Eq. (\ref{3})), both in the bare as well as in the renormalized theory
(they don't have to be the same positive integer). Clearly, this is
possible for arbitrary integers and colour factors, only if the series,
on the right hand side of Eq. (\ref{31}) terminates after the second
term, namely, only if there is no contribution in Eq. (\ref{31})
beyond one loop. Our proof explicitly verifies that this expectation
is, indeed, true. However, it is important to recognize that our proof
uses constraints coming only from the behaviour under {\it small gauge}
invariance (and, of course, analyticity) much like the proof in the
Abelian case. It is worth remarking here that, in a recent paper
{\cite{20}}, it
has been argued, using a generalization of the method of holomorphy
due to Seiberg {\cite{21}}, that in a Yang-Mills theory
interacting with matter 
fields, {\it without a tree level Chern-Simons term}, there is no
higher loop renormalization of the induced Chern-Simons
coefficient. Our result, for the case with a tree level Chern-Simons
term, is not covered by this analysis (as the authors of
ref. {\cite{20}}
specifically point out) and, in fact, this case is physically more
meaningful since, without a tree level Chern-Simons term, a loop
expansion of the theory may not exist because of severe infrared
divergences. In such a case, general formal arguments may be
invalidated by the infrared divergences of the perturbation theory. 

\section{Pure Chern-Simons theory:}

In this section, we will study in detail the pure Chern-Simons theory
{\cite{22}}
(otherwise also known as the $\epsilon$-theory
{\cite{7}}) in the infrared safe
axial gauge and show that it is a free theory. The pure Chern-Simons
theory can be obtained from the Lagrangian density in Eq. (\ref{1})
(or (\ref{2})) by dropping the Yang-Mills term (namely, it is the theory
in the $m\rightarrow \infty$ limit). It is well known that, in the
Landau gauge, this theory is invariant under a vector supersymmetry
{\cite{23}},
in addition to the usual BRST symmetry of Eq. (\ref{4}). Namely, the
Lagrangian density 
\begin{equation}
{\cal L} = {m\over
2}\,\epsilon^{\mu\nu\lambda}A_{\mu}^{a}(\partial_{\nu}A_{\lambda}^{a}
+ {g\over 3} f^{abc}A_{\nu}^{b}A_{\lambda}^{c}) -
F^{a}(\partial_{\mu}A^{\mu a}) +
\partial^{\mu}\overline{c}^{a}(D_{\mu}c^{a})
\end{equation}
is, of course, invariant under the BRST transformations of
Eq. (\ref{4}), but it is also invariant under the transformations,
\begin{eqnarray}
\delta A_{\mu}^{a} & = & \epsilon_{\mu\nu\lambda}
\epsilon^{\nu}\partial^{\lambda}c^{a} \nonumber\\
\noalign{\kern 4pt}%
\delta c^{a} & = & 0\nonumber\\
\noalign{\kern 4pt}%
\delta \overline{c}^{a} & = & \epsilon^{\mu}A_{\mu}^{a}\nonumber\\
\noalign{\kern 4pt}%
\delta F^{a} & = & \epsilon^{\mu}(D_{\mu}c^{a})
\end{eqnarray}
Here $\epsilon^{\mu}$ is a constant vector parameter of the
transformations and is anti-commuting in nature. Furthermore, the
generators of these transformations satisfy a supersymmetry algebra,
unlike the BRST charges which are nilpotent.

Let us next show that this supersymmetry is not particular to the
Landau gauge only. It is easy to see that there is a vector
supersymmetry in the axial gauge as well. Thus, the Lagrangian density
\begin{equation}
{\cal L} = {m\over 2}\,\epsilon^{\mu\nu\lambda} A_{\mu}^{a}
(\partial_{\nu}A_{\lambda}^{a} + {g\over 3} f^{abc}
A_{\nu}^{b}A_{\lambda}^{c}) - F^{a}(n^{\mu}A_{\mu}^{a}) -
\overline{c}^{a}n^{\mu}(D_{\mu}c^{a}) 
\end{equation}
is invariant under the BRST transformations of Eq. (\ref{4}) as well
as the transformations
\begin{eqnarray}
\delta A_{\mu}^{a} & = & \epsilon_{\mu\nu\lambda}
\epsilon^{\nu}n^{\lambda}c^{a} \nonumber\\
\noalign{\kern 4pt}%
\delta c^{a} & = & 0\nonumber\\
\noalign{\kern 4pt}%
\delta \overline{c}^{a} & = & -\epsilon^{\mu}A_{\mu}^{a}\nonumber\\
\noalign{\kern 4pt}%
\delta F^{a} & = & \epsilon^{\mu}\partial_{\mu}c^{a}\label{wi}
\end{eqnarray}
In fact, it is quite easy to check that, in any linear, homogeneous
infrared safe gauge, the theory develops an invariance under a vector
supersymmetry. 

Let us analyze the pure Chern-Simons theory in the axial gauge. In
such a case, there is the usual Ward identities following from the
BRST invariance of Eq. (\ref{4}). And, as we have noted earlier, the
structure of the theory leads to the fact that, in the axial gauge,
there is no wave function or vertex renormalization for the
ghosts. Let us note now that the new vector supersymmetry will also
lead to a Ward identity, further restricting the amplitudes. The master
identity, following from the invariance of the theory under
Eq. (\ref{wi}) takes the form (We note here that the derivation of
this identity is much simpler than the usual Ward identities because
the transformations in Eq. (\ref{wi}) are, in fact, linear
and, consequently, we do not need additional sources in the Lagrangian
density.)
\begin{equation}
\int d^{3}x\left(\epsilon_{\mu\nu\lambda}\,{\delta\Gamma\over\delta
A_{\nu}^{a}(x)} c^{a}(x)n^{\lambda} - {\delta\Gamma\over \delta
F^{a}(x)} \partial_{\mu}c^{a}(x) + A_{\mu}^{a}(x) {\delta\Gamma\over
\delta\overline{c}^{a}(x)}\right) = 0\label{wii}
\end{equation}
This, indeed, constrains the theory enormously. Combining with the
facts that the $F^{a}A_{\mu}^{a}$ vertex, the ghost two point vertex
and the ghost interaction vertices are not renormalized, it
immediately leads us to the result that the two point and the three
point functions for the gauge fields are not renormalized
either. For example, taking derivative of Eq. (\ref{wii}) with respect to
${\delta^{2}\over \delta A_{\mu}^{a}\delta c^{b}}$ (index $\mu$ being
summed) and setting all
fields to zero, we obtain in momentum space
\begin{equation}
\epsilon_{\mu\nu\lambda}n^{\lambda}{\delta^{2}\Gamma\over \delta
A_{\mu}^{a}(-p)\delta A_{\nu}^{b}(p)} + i
p_{\mu}{\delta^{2}\Gamma\over \delta A_{\mu}^{a}(-p)\delta F^{b}(p)} -
3{\delta^{2}\Gamma\over \delta\overline{c}^{a}(-p)\delta c^{b}(p)} = 0
\end{equation}
This immediately leads to $Z_{m}=1$. Similarly, taking one higher
derivative, it is easy to show that the parity violating three point
vertex function is not renormalized either.
In other words, the pure Chern-Simons theory is a free theory. 
%%% rev
These conclusions are valid provided one regularizes the theory such
that all symmetries, including the vector supersymmetry, are maintained.
%%%
We would like to note here that such a conclusion was reached earlier
from different points of view {\cite{7,refA,refB,24}}.
%%rev
In particular, in reference \cite{refA}, it was shown through a
perturbative calculation in the pure Chern-Simons theory, that the
complete effective action in axial gauges is the three level action,
for certain classes of gauge invariant regulators. 
Here we have derived this result from purely algebraic considerations.

\section{Conclusion:}

In this paper, we have studied in detail the question of higher order
corrections to the Chern-Simons coefficient in a
Yang-Mills-Chern-Simons theory. We have shown that the Chern-Simons
coefficient is, in general, a gauge dependent quantity. However, it
takes on a physical significance in the axial gauge. Using, i) the Ward
identities of the theory and, ii) the analyticity of the amplitudes in
the momentum variables, we have shown that, in the axial gauge, the
Chern-Simons coefficient does not receive any quantum correction
beyond one loop. This allows us to deduce that the ratio ${4\pi m\over
g^{2}}$, in a non-Abelian theory, is not renormalized beyond one loop,
in  any infrared safe
gauge. This, therefore, represents the generalization of the
Coleman-Hill result to a non-Abelian theory. Various other interesting
properties of the theory are also discussed. 

\bigskip 

This work was supported in part by U.S. Dept. Energy Grant DE-FG
02-91ER40685  as well as by CNPq, Brazil. 

\bigskip

\providecommand{\href}[2]{#2}\begingroup\raggedright
\endgroup

\end{document}